\documentclass[prb,twocolumn,aps,amssymb,footinbib,showpacs]{revtex4}
\usepackage{amssymb}
\usepackage{graphicx}
\usepackage{amsmath}
\usepackage{times}
\usepackage{color}
\usepackage{subfigure}
\usepackage{setspace}
\usepackage{bm}

\begin{document}

\newcommand {\beq} {\begin{equation}}
\newcommand {\eeq} {\end{equation}}
\newcommand {\bqa} {\begin{eqnarray}}
\newcommand {\eqa} {\end{eqnarray}}
\newcommand {\no} {\nonumber}
\newcommand {\ep} {\ensuremath{\epsilon}}
\newcommand {\ms} {\medskip}
\newcommand {\bs} {\bigskip}
\newcommand{\rr} {\ensuremath{{\bf r}}}
\newcommand{\nbr} {\ensuremath{\langle ij \rangle}}
\newcommand{\nbrr} {\ensuremath{\langle kl \rangle}}
\newcommand{\ncap} {\ensuremath{\hat{n}}}
\newcommand{\can} {\ensuremath{{\cal S}}}
\newcommand{\vep}{\ensuremath{\varepsilon}}
\newcommand {\pcl} {\ensuremath{\phi_{{\rm c}}}}
\newcommand {\pcla} {\ensuremath{\phi^\ast_{{\rm c}}}}
\newcommand {\kk} {\ensuremath{{\bf{k}}}}
\newcommand {\la} {\ensuremath{\Lambda}}
\newcommand {\phl} {\ensuremath{\phi^{<}}}
\newcommand {\phg} {\ensuremath{\phi^{>}}}
\newcommand {\phga} {\ensuremath{\phi^{\ast>}}}
\newcommand {\pc} {\ensuremath{\phi_{c}}}
\newcommand {\pq} {\ensuremath{\phi_{q}}}
\newcommand {\pca} {\ensuremath{\phi^\ast_{c}}}
\newcommand {\pqa} {\ensuremath{\phi^\ast_{q}}}
\newcommand {\pqla} {\ensuremath{\phi^{<\ast}_{q}}}
\newcommand {\pcga} {\ensuremath{\phi^{>\ast}_{c}}}
\newcommand {\pqga} {\ensuremath{\phi^{>\ast}_{q}}}
\newcommand {\pql} {\ensuremath{\phi^{<}_{q}}}
\newcommand {\pcg} {\ensuremath{\phi^{>}_{c}}}
\newcommand {\pqg} {\ensuremath{\phi^{>}_{q}}}

\title{A perturbative renormalization group approach to driven quantum systems}
\date{\today}

\author{ Sangita De Sarkar$^1$, Rajdeep Sensarma$^2$, and K. Sengupta$^1$}
 \affiliation{ 1.~~ Theoretical Physics Department, Indian Association for the Cultivation of Science, Kolkata 700032, India\\
 2.~~Department of Theoretical Physics, Tata Institute of Fundamental Research, Mumbai 400005, India}

\begin{abstract}
We use a perturbative momentum shell renormalization group (RG)
approach to study the properties of a driven quantum system at zero
temperature. To illustrate the technique, we consider a bosonic
$\phi^4$ theory with an arbitrary time dependent interaction
parameter $\lambda(t)=\lambda  f(\omega_0 t)$, where $\omega_0$ is
the drive frequency and derive the RG equations for the system using
a Keldysh diagrammatic technique. We show that the scaling of
$\omega_0$ is analogous to that of temperature for a system in
thermal equilibrium and its presence provides a cutoff scale for the
RG flow. We analyze the resultant RG equations, derive an analytical
condition for such a drive to take the system out of the gaussian
regime, and show that the onset of the non-gaussian regime occurs
concomitantly with appearance of non-perturbative mode coupling
terms in the effective action of the system. We supplement the
above-mentioned results by obtaining them from equations of motions
of the bosons and discuss their significance for systems near
critical points described by time-dependent Landau-Ginzburg
theories.
\end{abstract}

\maketitle

\section{Introduction}
\label{intro}

The flow of coupling parameters of an action (or equivalently a
Hamiltonian) under renormalization group (RG) transformations plays
a central role in understanding the low-energy properties of the
system described by the action. This also provides us with a way of
understanding the concept of universality which refers to the fact
that systems described by different microscopic Hamiltonians show
identical scale independent low-energy behavior specially near
critical points. The microscopic action describing a quantum system
may have many parameters and thus be complicated; however, many of
these parameters might turn out to be irrelevant for phenomena
involving low-energy or low-momenta. This leads to a simpler
effective action with fewer parameters which describes the
low-energy properties of the system. This concept is central to
understanding the validity of attempts to explain, for example, the
low-temperature experimental data of a quantum system based on
simple model actions. The procedure for obtaining such an effective
action is well known for equilibrium systems \cite{rg1}. For weakly
interacting systems, where the interaction term in the action can be
treated perturbatively, this can be done analytically; the analysis
of the resultant RG equations provide useful information of the
coupling parameters, and hence the effective action, of the system
at an arbitrary length scale\cite{rg2}.

In recent years, there has been a lot of theoretical and
experimental interest in studying intrinsic quantum dynamics of
strongly interacting many-body systems \cite{rev1}. This interest is
largely due to recent experimental realization of such isolated
quantum systems in form of ultracold atoms in optical lattices
\cite{blochrev} which act as perfect test bed for such dynamics. The
suitability of these systems in this regard originates from their
near-perfect isolation from the surrounding which leads to long time
scale over which quantum dynamics can be observed. However, we note
here that more recently pump-probe experiments have also started to
probe non equilibrium dynamics in the context of standard materials
based condensed matter systems \cite{ppexp}.

The equilibrium properties of ultracold atoms are generically
described by using simple model Hamiltonians such as the
Bose-Hubbard model \cite{bloch1} or Ising model in transverse and
longitudinal fields \cite{greiner1}; indeed, one of the main
interest in ultracold atom systems stems from their role as
emulators of well-studied models of quantum statistical mechanics.
However, the description of a complicated coupled atom-laser system
in terms of simple quantum models at low energies invariably relies
on the concept of universality. This procedure is conceptually
justified by invoking standard RG arguments in equilibrium which
leads to an effective action using the following steps. First, one
imposes a ultraviolet momentum cutoff $\Lambda$ , which, in a
typical condensed matter system, is roughly the inverse of the
lattice spacing. Second, this cutoff is lowered from $\Lambda$ to
$\Lambda -d\Lambda$ and the field modes within the momentum shell
$\Lambda$ and $\Lambda - d\Lambda$ is integrated out perturbatively
(in the simplest case to one-loop order in interaction) to obtain an
effective action describing the field modes below the cutoff
$\Lambda-d\Lambda$. Next, the momentum and the frequency in the
action are rescaled appropriately so as to offset the change in the
cutoff. Finally, one reads out the change in parameters of the
action due to the set of transformations described above (rescaling
and integrating out the field modes within the momentum shell) and
obtains the resultant RG flow equation for the parameters of the
action. Such a flow leads to either increase (relevant) or decrease
(irrelevant) of an Hamiltonian parameter; the low-energy effective
Hamiltonian is thus determined by only the relevant parameters which
leads to universality.

However, a well-defined RG procedure which can justify universality
in the long-time behavior of a generic out-of-equilibrium system is
not yet available in the literature. In fact, one of the central
questions in this field concerns the applicability of universality
in a driven quantum system for an arbitrary drive protocol. This
question has been partially addressed in a recent work studying the
role of a periodic potential in the time evolution following a
sudden quench of interaction parameter of an one-dimensional
Luttinger liquid \cite{mitra1}. Such an interaction is known to be
irrelevant for equilibrium situation; in contrast, Ref.\
\onlinecite{mitra1} found that such a term can play important role
in generation of dissipation and eventual thermalization of such a
system and can therefore not be neglected as irrelevant during
evolution after a quench. Similar studies have been carried out for
other non-equilibrium low-dimensional driven systems using
generalizations of Hamiltonian flow methods \cite{wegner1,kehrein1}.
However, the situation for higher dimensional systems and for
finite-rate protocols is presently far from clear.

In this work, we consider a driven bosonic system which is described
by a $\phi^4$ field theory with the action $S= S_0 + S_1$
\begin{eqnarray}
S_0 &=& \int \, \frac{d^dk d\omega}{(2\pi)^{d+1}}\, \phi^{\ast}
({\bf k},\omega) (g(\omega)-v_0^2
|{\bf k}|^2-r) \phi ({\bf k},\omega) \nonumber\\
S_1 &=& -\int d^d x dt \lambda(t) |\phi({\bf x},t)|^4,  \label{ac1}
\end{eqnarray}
where $g(\omega)$ depends on the dynamical critical exponent $z$  of
the theory and takes values $\omega(\omega^2)$ for $z=2(1)$, $v_0$
is the velocity and $r$ is the square of the mass of the bosons and
$\lambda(t)=\lambda f(\omega_0 t)$ is the time dependent interaction
parameter, $f$ is an arbitrary function, and $\omega_0$ is the drive
frequency. We carry out a perturbative momentum-shell RG analysis of
this action which leads to the following results. First, we show
that the drive frequency $\omega_0$ scales in the same manner as
temperature in equilibrium systems \cite{rg1} and provides a new
cutoff scale for the RG flow. Second, by analyzing the RG equations
for $r$ and $\lambda$, we identify two regimes for such driven
systems; in the first regime the drive can be treated perturbatively
and the concept of universality holds similar to that in equilibrium
situation while in the second, the drive dominates the physics and
determines the cutoff scale (similar to temperature in an
equilibrium system) for RG flow. We provide a criterion for
crossover between these two regimes for arbitrary drive protocol.
Third, we show that in the second regime, the presence of the drive
may take the system out of the gaussian regime (where the
interaction term of the effective low-energy action can be treated
perturbatively). At the onset of this non-gaussian regime, the
coupling between the different field modes due to the interaction
becomes comparable to the mass term in the action. We provide an
analytical condition involving $r$, $\lambda$ and $\omega_0$ for
this phenomenon to take place and discuss its relation to the onset
of dynamical transition studied in Ref.\ \onlinecite{pol1}. Finally,
we supplement the above-mentioned results by obtaining their analog
from an equation of motion method and discuss the relevance of our
analysis for near-critical systems described by time-dependent
Landau-Ginzburg theories.

The plan of the rest of the paper is as follow. In Sec.\
\ref{keldyshsec}, we analyze $S$ (Eq.\ \ref{ac1}) using a Keldysh
formalism and obtain the RG equations for its parameters. This is
followed by analysis of these equations in Sec. \ref{rganalysis}
where we obtain analytical condition for the onset of the
non-Gaussian regime. Next, we analyze the equation of motion for the
bosons in Sec.\ \ref{gpan}. We discuss our main results and conclude
in Sec.\ \ref{diss} and provide some detail of the calculations in
the appendix.

\section{Computation of RG equations}
\label{keldyshsec}

\begin{figure}
\rotatebox{0}{\includegraphics*[width=0.8 \linewidth]{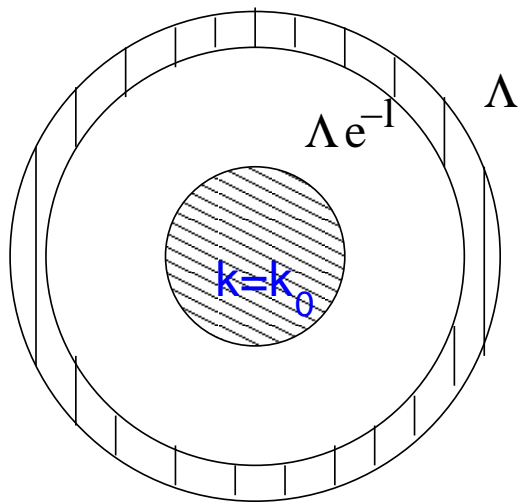}}
\caption{Schematic representation of momentum space picture for the
driven $\phi^4$ theory. The dashed circle with radius $\Lambda_2=
\Lambda \exp(-\ell_2)$ represents the set of states which
participates in the dynamics. The cutoff is lowered from $\Lambda$
to $\Lambda \exp(-\ell)$ and the momentum states within this shell
is integrated out. The RG flow stops when the two circles touch each
other at $\ell=\ell_2$; see text for details.} \label{fig1}
\end{figure}

In this section, we analyze $S$ (Eq.\ \ref{ac1}) using Keldysh
technique which is ideally suited for handling out-of-equilibrium
quantum systems \cite{kamenev1}. To this end, we follow standard
procedure to introduce the fields $\phi_+({\bf k},\omega)$ and
$\phi_{-}({\bf k}, \omega)$ living on the forward and backward time
contours. In terms of these fields, the zero temperature partition
function for a system of interacting bosons can be written as
\begin{eqnarray}
Z &=& \int {\mathcal D} \phi_+ {\mathcal D} \phi_- e^{i (S_+[\phi_+]
- S_-[\phi_-])}, \label{part1}
\end{eqnarray}
where $S[\phi]$ is given by Eq.\ \ref{ac1}. Next, for computational
convenience, we define classical and quantum components of the
bosonic fields $\phi$ as
\begin{eqnarray}
\phi_{{\rm c}({\rm q})} &=& \left(\phi_+ +(-) \phi_-\right)/2
\label{def1}
\end{eqnarray}
and write the partition function as
\begin{eqnarray}
Z &=& \int {\mathcal D} \phi_{\rm c} {\mathcal D} \phi_{\rm q} e^{i
S'[\phi_{\rm c},\phi_{\rm q}]}, \label{part2}
\end{eqnarray}
where the action $S'=S'_0 +S'_1$ is given by
\begin{eqnarray}
S^{'}_0 &=& 2 \int \frac{d^dk d\omega}{(2\pi)^{d+1}}\, \phi^{\ast}
({\bf k},\omega) (g(\omega)-v_F^2 k^2-r) \sigma_x \phi
({\bf k},\omega) \nonumber\\
S'_1 &=& -4 \int d^dx dt \, \lambda(t) \left[ \pcla({\bf
x},t)\pqa({\bf x},t) \left\{ \pcl({\bf x},t)\pcl({\bf x},t)  \right.
\right. \nonumber\\
&& \left. \left. + \pq({\bf x},t) \pq({\bf x},t) \right\}  +\rm
{h.c} \right]. \label{kelac1}
\end{eqnarray}
Here $\phi^{\ast}= (\phi_{\rm c}^{\ast},\phi_{\rm q}^{\ast})$ is the
two component bosonic field and $\sigma_x$ is the Pauli matrix
acting in ${\rm c-q}$ space.

To analyze this action using perturbative RG, we first rewrite
$S'_1$ in momentum-frequency space. To this end, we define a
dimensionless kernel
\begin{eqnarray}
K(\alpha) = \int_{-\infty}^{\infty} dy  f(y) \exp(i\alpha y)
\label{kexp}
\end{eqnarray}
and rewrite $\int_{-\infty}^{\infty} dt f(\omega_0 t) \exp(i \omega
t) = K(\omega/\omega_0)/\omega_0$. In terms of this dimensionless
kernel, one can write
\begin{eqnarray}
S'_1 &=& -4 \prod_{i=1}^3 \prod_{j=1}^4 \int \frac{ d^d k_i
d\omega_j}{(2\pi)^{3d+4} \omega_0} \lambda K(\omega/\omega_0) \left[
\pcla({\bf k_1},\omega_1) \right. \nonumber\\
&& \left. \times \pqa({\bf k_2},\omega_2) \left\{ \pcl({\bf
k_3},\omega_3)\pcl({\bf k_1+k_2-k_3},\omega_4)  \right.
\right. \nonumber\\
&& \left. \left. + \pq({\bf k_3},\omega_3) \pq({\bf
k_1+k_2-k_3},\omega_4) \right\} +\rm {h.c} \right], \label{intac1}
\end{eqnarray}
where $\omega= \omega_1+\omega_2-\omega_3-\omega_4$. We note that
the physical significance of $K$ is that it encodes the manner in
which different modes are coupled by the interaction. For example,
for a periodic drive with $f(\omega_0 t)= a + b \cos(\omega_0 t)$,
one can show that $K(\omega/\omega_0) = \sum_{n} \alpha_n
\delta(\omega/\omega_0 -n)$ with $\alpha_0=a$,
$\alpha_1=\alpha_{-1}=b/2$ and $\alpha_{|n|>1}=0$. It will be shown
that such an interaction leads to coupling between field modes
$\phi({\bf k},\omega)$ and $\phi({\bf k},\omega - n\omega_0)$ with
amplitude $\alpha_n$. In contrast, for a gaussian drive profile with
$f(\omega_0 t) \simeq \exp(-\omega_0^2 t^2/2)$, one finds
$K(\omega/\omega_0) \simeq \exp[-\omega^2/(2\omega_0^2)]$; here, as
we shall derive subsequently, any two field modes with frequencies
$\omega$ and $\omega'$ are coupled to each other with a strength
$\sim \exp[-(\omega-\omega')^2/(2\omega_0^2)]$. We would like to
stress that values of $K(\omega/\omega_0)$ (or $\alpha_n$ for
periodic drive) depends on the drive protocol. In what follows we
shall keep $f(\omega_0 t)$ arbitrary except for the requirement that
$\int_{-\infty}^{\infty} dy  f(y) \exp(i\alpha y)$ is well defined.
We also note that we envisage a situation in which the drive decays
to zero with a characteristic time scale $T_0$. The presence of
$T_0^{-1}$ changes the expressions for $K(\omega)$ but, as we shall
see, do not influence the RG flow otherwise provided $T_0^{-1} \le
\omega_0$. For example, for periodic drives with a gaussian decaying
profile $f(\omega_0 t) =(a+b \cos(\omega_0 t) \exp[-t^2/2T_0^2]$,
one has $K(\omega/\omega_0) \simeq a \exp(-\omega^2 T_0^2/2) + b/2
(\exp[-(\omega-\omega_0)^2 T_0^2/2] + \exp[-(\omega+\omega_0)^2
T_0^2/2])$ which reduces to earlier derived results for large $T_0$.
However, having a finite $T_0$ is important in the present case
since in the absence of a reservoir, for $T_0 \to \infty$, the
system will heat up indefinitely. In this case, the system reaches
the infinite temperature fixed point where the low-energy effective
action looses its meaning.

Next, we present our rationale for feasibility of a RG analysis of
the driven system. We consider the system to be in the ground state
of $S'$ at the start of drive labeled by a momenta ${\bf k_0}$. The
central assumption of the RG analysis that follows is that for any
generic action, there will be a finite set of states in the Hilbert
space around ${\bf k_0}$, as schematically shown in Fig.\
\ref{fig1}, which will actively participate in the dynamics. The
number of such states depends on the drive frequencies and
amplitude. The other states in the Hilbert space do not participate
in the dynamics and may thus be systematically integrated out to
obtain an effective action for the system in terms of the active
modes. In what follows, we are going to implement this procedure. In
doing so, we follow the convention of imposing a finite momentum
cutoff leaving the frequency cutoff to infinity \cite{rg1}. The
first step of the RG transformation is scaling which constitutes
lowering of the momentum cut-off $\la$ to $\la e^{-l}$ leading to
the slow and the fast field modes given by \bqa \phi(k)=\left\{
\begin{array}{l}
\phi^< \quad \ 0<k<\la e^{-l}\\
\phi^> \quad \ \la e^{-l}<k<\la
\end{array}
\right. \eqa In perturbative RG, the fast modes are eliminated by
integrating them out perturbatively keeping only one-loop terms in
the interaction $\lambda$, followed by a standard rescaling of the
resultant effective action. Such an elimination of the fast modes
leads to
\bqa S^\prime(\phi^<) &=& S_0(\phi^<)+\langle
S_1(\phi^<,\phi^>)\rangle_{S^>_{0}}\no\\&&+\frac{1}{2}\left(\langle
S_1^2\rangle_{S^>_{0}}
-\langle S_1\rangle^2_{S^>_{0}}\right)+....\no\\
&=& S_0(\phi^<)+S_1(\phi^<)+S_2(\phi^<), \no \eqa
where $S_2$ results from one-loop corrections from the interaction
terms and is derived in Sec.\ \ref{appa}.

We first consider scaling of $S_0$ and $S_1$. To this end, we follow
the standard procedure of rescaling, namely, $k \rightarrow k
\exp(\ell)$, $\phi^< \rightarrow  \phi^< \exp(-\alpha)$, and $\omega
\rightarrow \omega \exp(z \ell)$. The invariance of $S_0$ under this
scaling demands $r\to r'= r \exp(2\ell)$ and $\alpha= (d+z+2)/2$
which fixed the scaling of the fields. The invariance of $S_1$ is
slightly more tricky; for this we note that $K$ is a dimensionless
function which does not scale under RG. Thus the invariance of $S_1$
requires $\lambda/\omega_0 \rightarrow (\lambda/\omega_0)
\exp[(\epsilon-z)\ell]$, where $\epsilon=4-d-z$. We choose the
simplest possible protocol independent solution (demanding that
$\omega_0 t$, being dimensionless, will remain invariant under
scaling) of this equation which is given by $\lambda \rightarrow
\lambda \exp(\epsilon \ell)$ and $\omega_0 \rightarrow \omega_0
\exp(z\ell)$. This leads to the tree level RG equations \bqa
\frac{dr(\ell)}{d \ell}&=&2r(\ell) \quad \frac{d\omega_0(\ell)}{d
\ell} = z\omega_0(\ell) \quad \frac{d\lambda(\ell)}{d \ell} =
\ep\lambda (\ell) \label{rgtree} \eqa within the initial condition
$r(0)=r$, $\lambda(0)=\lambda$ and $\omega_0(0)=\omega_0$. From Eq.\
\ref{rgtree}, we note that the drive frequency $\omega_0$ scales as
$\omega_0(\ell) = \omega_0 \exp(z \ell)$ showing that it is relevant
under RG. The scaling of $\omega_0$ is reminiscent of the scaling of
physical temperature in equilibrium systems \cite{rg2} which is
known to scale as $ T(\ell)= T \exp(z \ell)$.

The full RG procedure which involves integrating out the fast modes
is worked out in details in the appendix. The resultant RG equations
are given by
\begin{eqnarray}
\frac{d r(\ell)}{d \ell} &=& 2r(\ell) + c_1 K(0) \lambda(\ell),
\quad \omega_0(\ell) = \omega_0 e^{z\ell}, \nonumber\\
\frac{d \lambda(\ell)}{d \ell} &=& \epsilon \lambda(\ell) - c_2 K(0)
\lambda^2(\ell), \label{rgeq1} \\
\frac{d r(\omega;\ell)}{d \ell} &=& c_1
K\left(\frac{\omega}{\omega_0}\right) \lambda(\ell), \quad \omega
\ne 0, \nonumber\\
\frac{d \lambda(\omega,\omega';\ell)}{d \ell} &=& - c_2
K\left(\frac{\omega}{\omega_0}\right)
K\left(\frac{\omega'}{\omega_0}\right) \lambda^2(\ell), \quad
\omega, \omega' \ne 0. \nonumber\
\end{eqnarray}
Here $c_1$ and $c_2$ are constants whose expressions are given in
the appendix, $\epsilon= 4-d-z$, and $r(\omega;\ell) \equiv
r(\omega)$ and $\lambda(\omega,\omega';\ell) \equiv \lambda(\omega
,\omega')$ are coefficients of terms generated in the action due to
integrating out the field modes within the shell $\Lambda$ and
$\Lambda \exp(-\ell)$. These terms are derived in the appendix and
are given by
\begin{eqnarray}
\delta S_1 &=& -2 \int \frac{d^d k d \omega_1
d\omega'}{(2\pi)^{d+2}\omega_0} \phi^{\ast <}({\bf k}, \omega_1)
r(\omega') \sigma_x \nonumber\\
&& \times\phi^<({\bf
k},\omega_1-\omega')  [1-\delta(\omega/\omega_0)] \nonumber\\
\delta S_2 &=& 4 \prod_{i=1}^3 \prod_{j=1}^4 \int \frac{ d^d k_i
d\omega_j d\omega d \omega'}{(2\pi)^{3d+6}\omega_0^2} \lambda
(\omega,\omega') \left[
\phi_c^{\ast <}({\bf k_1},\omega_1) \right. \nonumber\\
&& \left. \times \phi_q^{\ast <}({\bf k_2},\omega_2) \left\{
\phi_c^<({\bf
k_3},\omega_3 -\omega) \right. \right. \nonumber\\
&& \left. \left. \times \phi_c^<({\bf k_1+k_2-k_3},\omega_4-\omega')
+ \phi_q^<({\bf k_3},\omega_3-\omega) \right. \right. \nonumber\\
&& \left. \left. \times  \phi_q^<({\bf
k_1+k_2-k_3},\omega_4-\omega')
\right\} +\rm {h.c} \right] \nonumber\\
&& \times [1-\delta(\omega/\omega_0)][1-\delta(\omega'/\omega_0)].
\label{intac2}
\end{eqnarray}
They are not present in the original action but are spontaneously
generated due to the RG flow. They represent quadratic and quartic
couplings between field modes with different frequencies due to the
time dependent drive. These terms keep track of the transfer of
energy between the field modes due to the drive and have no analog
in equilibrium RG. We also note that for periodic drive, $K(0)$ and
$r(\omega)$ should be carefully defined since $K(\omega/\omega_0)$
has supports on discreet points where $\omega/\omega_0=n$. As shown
in the appendix, in this case one obtains
\begin{eqnarray}
\frac{d r(\ell)}{d \ell} &=& 2r(\ell) + c_1 \alpha_0 \lambda(\ell),
\quad \omega_0(\ell) = \omega_0 e^{z\ell}, \nonumber\\
\frac{d \lambda(\ell)}{d \ell} &=& \epsilon \lambda(\ell) - c_2
\alpha_0 \lambda^2(\ell), \label{rgeq2} \\
\frac{d r_n(\ell)}{d \ell} &=& c_1 \alpha_n \lambda(\ell), \quad n
\ne 0, \nonumber\\
\frac{d \lambda_{mn}(\ell)}{d \ell} &=& - c_2 \alpha_n \alpha_m
\lambda^2(\ell), \quad m,n \ne 0. \nonumber\
\end{eqnarray}
with the additional terms generated in the action being given by
\begin{eqnarray}
\delta S'_1 &=& -2 \sum_{n\ne 0} \int \frac{d^d k d \omega_1
}{(2\pi)^{d+1}} \phi^{\ast <}({\bf k}, \omega_1) r_n \sigma_x
\nonumber\\
&& \times \phi^<({\bf k},\omega_1-n \omega_0)\nonumber\\
\delta S'_2 &=& 4 \prod_{i=1}^3 \prod_{j=1}^4 \sum_{m,n\ne 0} \int
\frac{ d^d k_i d\omega_j }{(2\pi)^{3d+4}} \lambda_{m n} \left[
\phi_c^{\ast <}({\bf k_1},\omega_1) \right. \nonumber\\
&& \left. \times \phi_q^{\ast <}({\bf k_2},\omega_2) \left\{
\phi_c^<({\bf
k_3},\omega_3 -m \omega_0) \right. \right. \nonumber\\
&& \left. \left. \times \phi_c^<({\bf k_1+k_2-k_3},\omega_1+\omega_2
-\omega_3-n \omega_0) \right. \right. \nonumber\\
&& \left. \left. + \phi_q^<({\bf k_3},\omega_3-m\omega_0) \right. \right. \\
&& \left. \left. \times  \phi_q^<({\bf
k_1+k_2-k_3},\omega_1+\omega_2-\omega_3-n\omega_0) \right\} +\rm
{h.c} \right]. \nonumber\ \label{intac3}
\end{eqnarray}
The RG equations derived here show that the drive frequency provides
a natural cutoff scale for the RG flow. We note that when
$\omega_0(\ell) \sim \Lambda(\ell)$, all the states in the Hilbert
space below the momentum cutoff participates in the dynamics and
hence one can not integrate out states any further. This cutoff
scale $\ell_2$ satisfies $\omega_0(\ell_2) = v_0 \Lambda(\ell_2)$,
and is given by
\begin{eqnarray}
\ell_2 &=& \frac{1}{z+1} \ln(\Lambda v_0/\omega_0). \label{cutoff1}
\end{eqnarray}
Note that there are  other cutoff scales in the problem stem from
the mass term $r$ since RG stops when the momentum cutoff reaches
inverse of the correlation length or when the interaction term grows
(for $\epsilon>0$) such that the perturbative RG analysis cease to
hold \cite{comment1}. We shall provide an explicit expression for
these scales in Sec.\ \ref{rganalysis}; here we simply note that for
large $\omega_0$, the RG flow stops at $\ell_2$.  Beyond
$\ell=\ell_2$, the property of the system is determined essentially
by the drive term and this regime has no analog in equilibrium RG.
We shall derive this explicitly in the next section.

Before moving on to the analysis of the RG equations, we note that
the one-loop correction terms to $r(\ell)$ and $\lambda(\ell)$
depends crucially on the driving protocol through $K(0)$ or
$\alpha_0$. This feature in turns leads to protocol dependent fixed
point structure for the RG equations. For example, drive protocols
with $K(0)[\alpha_0]=0$, the equations for $r(\ell)$ and
$\lambda(\ell)$ do not have a Wilson-Fisher fixed point for relevant
interactions ($\epsilon>0$); only the Gaussian fixed point exists in
this case.

\section{Analysis of RG equations}
\label{rganalysis}

The solutions of the RG equations (Eq.\ \ref{rgeq1}) depend
crucially on the relevance/irrelevance of the interaction. We begin
with the case when the interaction term is irrelevant, {\it i.e.},
$d+z > 4$. In this case since $\epsilon < 0$, it is possible to
ignore the second term in the right side of the RG equation for
$\lambda(\ell)$. Denoting $r$ and $\lambda$ to be the bare values of
$r(\ell)$ and $\lambda(\ell)$ and scaling all frequencies (momenta)
in units $\Lambda v_0 (\Lambda)$, we get
\begin{eqnarray}
r(\ell) &=&  r_{\rm eff} e^{2\ell} - c_1 K(0) \lambda e^{\epsilon \ell}/(d+z-2) \nonumber\\
\lambda(\ell) &=& \lambda e^{\epsilon \ell} \quad \omega_0(\ell)=
\omega_0 e^{z\ell} \nonumber\\
r(\omega;\ell) &=& -c_1 K(\omega/\omega_0) \lambda e^{\epsilon
\ell}/|\epsilon|  \nonumber\\
\lambda(\omega,\omega'; \ell) &=& c_2 K(\omega/\omega_0)
K(\omega'/\omega_0) \lambda^2 e^{2 \epsilon
\ell}/2|\epsilon|,\label{rgsolirr}
\end{eqnarray}
where the effective mass $r_{\rm eff} = r +c_1 K(0) \lambda
/(d+z-2)$. For periodic drive, the solution of the RG equations can
be easily read off from Eq.\ \ref{rgsolirr} by replacing $K(0) \to
\alpha_0$, $r(\omega;\ell) \to r_n(\ell)$, $K(\omega/\omega_0) \to
\alpha_n$, $\lambda(\omega,\omega'; \ell) \to \lambda_{mn}$, and
$K(\omega'/\omega_0) \to \alpha_m$.

To analyze Eq.\ \ref{rgsolirr}, we note that the RG flow stops when
the momentum cutoff reaches the cutoff scale set by the drive
frequency $\omega_0$ or when it reaches the inverse of the
correlation length. The former occurs at $\ell=\ell_2 =
\ln(1/\omega_0)/(z+1)$ while the latter happens at a RG time
$\ell_1$ for which $ \Lambda(\ell_1)v_0 = r(\ell_1)$. This leads to
$\ell_1 \simeq \ln(1/r_{\rm eff})/2$. With these two scales, there
are two distinct regimes. In the first regime, $\ell_1 \le \ell_2$,
so that
\begin{eqnarray}
r_{\rm eff} \ge \omega_0^{2/(z+1)}, \label{l1l2cond}
\end{eqnarray}
and the RG stops when the momentum cutoff reaches the inverse
correlation length. In this regime $r_{\rm eff} \simeq 1$, and the
drive frequency remain small compared to $r_{\rm eff}$ if
$\omega_0(\ell_1) \ll 1$ which leads to the condition
\begin{eqnarray}
\omega_0 \ll r_{\rm eff}^{z/2}. \label{pertcond}
\end{eqnarray}
We note that if the condition given by Eq.\ \ref{pertcond} is
satisfied, then the drive can be treated perturbatively; this
condition becomes analogous to the condition for the existence of a
perturbative quantum regime in equilibrium systems where the role of
$\omega_0$ is played by the temperature \cite{rg2}. In this
perturbative regime, when the RG flow stops, $\lambda(\ell_1) =
\lambda r_{\rm eff}^{|\epsilon|/2}$ and is thus small provided
$\lambda/r_{\rm eff} \ll 1$. Also all higher powers of interaction
remains small and can therefore be ignored; thus we conclude that
the universality of the driven system remains qualitatively similar
to that in equilibrium situation in this regime.

\begin{figure}
\rotatebox{0}{\includegraphics*[width=\linewidth]{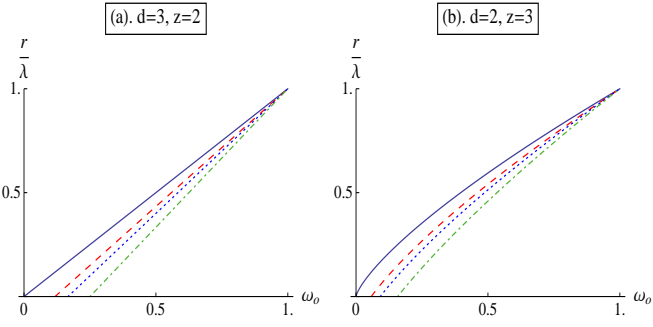}}
\caption{Plot of the $r/\lambda$ in case of irrelevant interaction
$\epsilon<0$ as a function of $\omega_0$ for a) $d=3$ and arbitrary
$z$ (left panel) and b)d=2 and z=3 (right panel) showing the border
between the drive induced non-gaussian and gaussian regimes for $c_1
K(0)=$ [or $c_1 \alpha_0=$] $0$ (blue solid line), $0.4$ (red dashed
line), $0.6$ (blue dotted line), and $1$ (green dash-dotted line).
Note that for large $c_1 K(0)$, the gaussian regime persists for any
non-zero $r/\lambda$ up to a critical drive frequency.} \label{fig2}
\end{figure}

In the second regime, RG flow stops at $\ell=\ell_2$ where
$\omega_0(\ell_2)=v_0 \Lambda(\ell_2)$. In this regime one finds
\begin{eqnarray}
r(\ell_2) &=& r_{\rm eff} \omega_0^{-2/(z+1)} - \frac{c_1 K(0)
\lambda}{d+z-2} \omega_0^{-\epsilon/(z+1)} \nonumber\\
\lambda(\ell_2) &=& \lambda \omega_0^{-\epsilon/(z+1)}. \label{rleq}
\end{eqnarray}
Thus the condition for non-gaussian behavior $\lambda(\ell_2) \ge
r(\ell_2)$ occurs when
\begin{eqnarray}
\lambda'/r_{\rm eff} \ge \omega_0^{(2-z-d)/(z+1)}, \label{ngcond}
\end{eqnarray}
where $\lambda'= \lambda[1+ c_1 K(0)/(z+d-2)]$. In this regime
$\ell_2 \le \ell_1$, and so we have $r_{\rm eff} \le
\omega_0^{2/(z+1)}$; thus a sufficient (but not necessary) condition
for violation of the Gaussian regime is given by
\begin{eqnarray}
\lambda' \ge \omega_0^{\epsilon/(z+1)}.  \label{suffcond}
\end{eqnarray}

\begin{figure}
\rotatebox{0}{\includegraphics*[width=\linewidth]{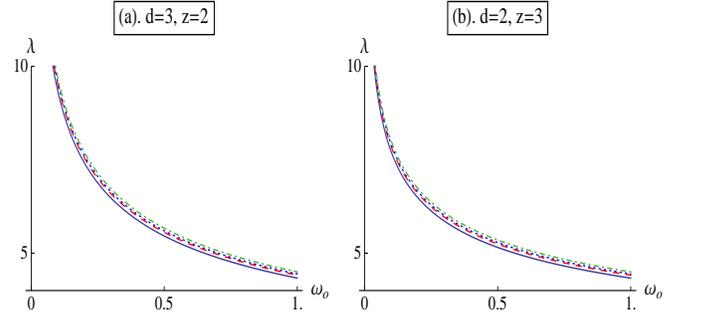}}
\caption{Plot of the $\lambda$ in case of irrelevant interaction as
a function of $\omega_0$ for a) $d=3$ and $z=2$ (left panel) and b)
$d=2$ and $z=3$ (right panel) showing the sufficient condition for
non-Gaussian regime for $c_1 K(0)=$ [or $c_1 \alpha_0=$] $0$ (blue
solid line), $0.4$ (red dashed line), $0.6$ (blue dotted line), and
$1$ (green dash-dotted line). } \label{fig3}
\end{figure}

Eqs.\ \ref{ngcond} and \ref{suffcond} constitute the central result
of this work. These equations show that the presence of a drive
frequency may stop the RG flow at a scale $\ell= \ell_2$. At this
scale, the system will exhibit non-gaussian behavior for a range of
$\lambda$ and $r_{\rm eff}$ for which Eq.\ \ref{ngcond} is
satisfied. As shown in Fig.\ \ref{fig2} and \ref{fig3}, the
condition for such a non-gaussian regime, for $d=3$ is given by
$\lambda' \omega_0 \ge r_{\rm eff}$ for any $z$. The sufficient
condition for $d=3$ and $z=2$ is given by $\lambda' \ge
\omega_0^{-1/3}$. In contrast for $d=2$, both the necessary and the
sufficient conditions depend on $z$; for $d=2$ and $z=3$, they are
given by $\lambda' \omega_0^{3/4} \ge r_{\rm eff}$ and $\lambda' \ge
\omega_0^{-1/4}$. Further, in this non-gaussian regime, one has
\begin{eqnarray}
r(\omega;\ell_2) &=& - c_1 K(\omega/\omega_0) \lambda
\omega_0^{-\epsilon/(z+1)}/|\epsilon| \\
\lambda(\omega,\omega'; \ell_2) &=& c_2 K(\omega/\omega_0)
K(\omega'/\omega_0) \lambda^2
\omega_0^{-2\epsilon/(z+1)}/(2|\epsilon|). \nonumber\
\label{rnterms}
\end{eqnarray}
This indicates that in the frequency range where $K(\omega/\omega_0)
\simeq 1$, $r(\omega,\ell_2)$ may become comparable to the mass
$r(\ell_2)$. Thus the onset of the non-gaussian regime indicates
that the drive may effectively transfer energy between different
modes. This is reminiscent of a dynamical energy delocalization
transition \cite{pol1} and we shall discuss this point further in
Sec.\ \ref{diss}. We also note that since $K(0)$ (or equivalently
$\alpha_0$ for periodic drive) depends on the protocol, the
condition of the onset of the non-gaussian regime may vary
drastically depending on the drive protocol. For larger values of
$K(0)$, one may have a regime where the non-gaussian behavior do not
show up for any finite $\lambda/r$ below a critical drive frequency.
This is reflected in Fig.\ \ref{fig2} where we sketch the condition
on $r/\lambda$ as a function of $\omega_0$ for several
representative values of $c_1K(0)$ or $c_1\alpha_0$. The
corresponding sufficiency condition for the onset of the of the
non-gaussian regime (Eq.\ \ref{suffcond}) is plotted in Fig.\
\ref{fig3}.

\begin{figure}
\rotatebox{0}{\includegraphics*[width=\linewidth]{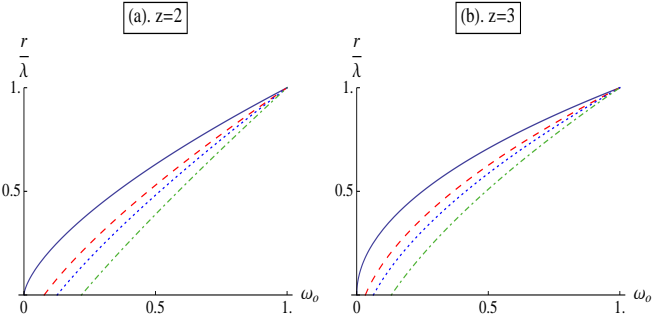}}
\caption{Plot of the $r/\lambda$ for marginal interaction as a
function of $\omega_0$ for a) $z=2$ (left panel) and b) $z=3$ (right
panel) showing the sufficient condition for non-Gaussian regime for
$c_1 K(0)=$ [or $c_1 \alpha_0=$] $0$ (blue solid line), $0.4$ (red
dashed line), $0.6$ (blue dotted line), and $1$ (green dash-dotted
line). For all plots we have chosen $c_2/c_1=0.5$.} \label{fig4}
\end{figure}

\begin{figure}
\rotatebox{0}{\includegraphics*[width=\linewidth]{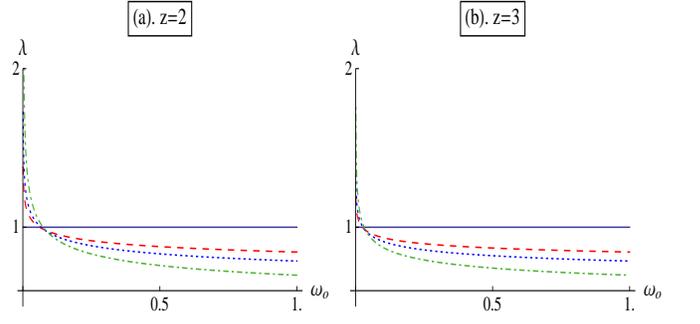}}
\caption{Plot of the $\lambda$ in case of marginal interaction as a
function of $\omega_0$ for a) $z=2$ (left panel) and b) $z=3$ (right
panel) showing the sufficient condition for non-Gaussian regime for
$c_1 K(0)=$ [or $c_1 \alpha_0=$] $0$ (violet solid line) $0.4$ (red
dashed line), $0.6$ (blue dotted line), and $1$ (green dash-dotted
line). All parameters are same as in Fig.\ \ref{fig4}.} \label{fig5}
\end{figure}

Next, we discuss the RG equation for the case of marginally
irrelevant interaction with $\epsilon=0$. For this, after some
straightforward algebra, one obtains the solution of the RG
equations to be
\begin{eqnarray}
r(\ell) &=& r'_{\rm eff} e^{2\ell} -c_1 K(0) \lambda
I(\ell;\lambda) \nonumber\\
\lambda(\ell) &=& \lambda (1+ c_2 K(0)\lambda \ell)^{-1} \\
r(\omega;\ell) &=& \frac{K(\omega/\omega_0)c_1}{K(0)c_2} \ln(1+c_2
K(0) \ell)\nonumber\\
\lambda(\omega,\omega';\ell) &=& \frac{K(\omega/\omega_0)
K(\omega'/\omega_0) \lambda}{K(0)[1+c_2 K(0) \lambda \ell]}, \quad
K(0)\ne 0, \nonumber\\
&=& c_2 K(\omega/\omega_0) K(\omega'/\omega_0) \lambda^2 \ell ,
\quad K(0)=0, \nonumber\ \label{rgmarginal}
\end{eqnarray}
where $r'_{\rm eff}$ and $I(\ell;\lambda)$ are given by
\begin{eqnarray}
r'_{\rm eff} &=& r + c_1 K(0) \lambda I(0; \lambda), \nonumber\\
I(\ell;\lambda) &=& \int^{\ell} d\ell' e^{-2\ell'}/(1+c_2 K(0)
\lambda \ell') \label{intdef}
\end{eqnarray}
The analysis of the RG equations proceed along the same line as the
one carried out for the earlier case. The RG flow stops at
$\ell=\ell_1$ if $r'_{\rm eff} \ge \omega_0^{2/(z+1)}$. In this
regime the drive can be treated perturbatively provided
$\omega_0/r_{\rm eff}^{' z/2} \ll 1$. In the other regime, where
$r'_{\rm eff} \le \omega_0^{2/(z+1)}$, the flow stops at
$\ell=\ell_2$. In this regime, one finds
\begin{eqnarray}
r(\ell_2) &=& r'_{\rm eff} \omega_0^{-2/(z+1)} \nonumber\\
&& -c_1 K(0) \lambda
I\left(-\ln(\omega_0)/(z+1);\lambda\right) \nonumber\\
\lambda(\ell_2) &=& \lambda [1- c_2 K(0)\lambda \ln(\omega_0)/(z+1)]^{-1} \\
r(\omega;\ell_2) &=& \frac{K(\omega/\omega_0)c_1}{K(0)c_2} \ln
\left[1-c_2
\lambda K(0) \frac{\ln(\omega_0)}{z+1}\right]\nonumber\\
\lambda(\omega,\omega';\ell_2) &=& \frac{K(\omega/\omega_0)
K(\omega'/\omega_0) \lambda}{1-c_2 K(0) \lambda \ln(\omega_0)/(z+1)}
\quad K(0)\ne 0, \nonumber\\
&=& c_2 K(\omega/\omega_0) K(\omega'/\omega_0) \lambda^2
\frac{\ln(\omega_0)}{z+1} , \quad K(0)=0, \nonumber\
\label{rgmarginal2}
\end{eqnarray}
The necessary and sufficient conditions for the violation of the
Gaussian regime $\lambda(\ell_2) \ge r'_{\rm eff}(\ell_2)$, is then
given by
\begin{eqnarray}
\lambda_m'/r'_{\rm eff} \ge \omega_0^{-2/(z+1)}, \quad {\rm and }
\quad \lambda_m' \ge 1 \label{margcond}
\end{eqnarray}
respectively, where $\lambda'_m = \lambda [1- c_2 K(0)
\ln(\omega_0)/(z+1) + c_1 K(0) I(-\ln(\omega_0)/(z+1);\lambda)]$.
Using Eq.\ \ref{margcond}, a plot of limiting values of $r/\lambda$
which separates the gaussian and the non-gaussian regimes vs the
drive frequency $\omega_0$ is shown in Fig.\ \ref{fig4}. These
relations become particularly simple for drive protocols for which
$K(0)=0$ (or equivalently $\alpha_0=0$). For these protocols, the
necessary condition for violation of the gaussian regime is given by
$\lambda/r \ge \omega_0^{-2/(z+1)}$. It is easy to see from Eq.\
\ref{rgmarginal2}, that in this limit $r(\omega;\ell_2) \sim
\lambda$ becomes comparable to $r(\ell_2)$ leading to onset of
transfer of energies between different field modes. The
corresponding sufficiency condition for the onset of the of the
non-gaussian regime is plotted in Fig.\ \ref{fig5}.

Finally, we consider the case for $\epsilon>0$. For theories with
$\epsilon>0$, the interaction grows with RG time. Consequently, in
equilibrium, the flow equations run towards the well-known
Wilson-Fisher fixed point at which $\lambda^{\ast}= \epsilon/c_2$
and $r^{\ast}= -c_1 \epsilon/(2c_2)$. For the driven system, the
position of the fixed point depends on $K(0)$; indeed, for $K(0)=0$,
the fixed point does not exists. The solution of the RG equations in
this case is straightforward and is given by
\begin{eqnarray}
r(\ell) &=& r_0 e^{2\ell}, \quad \lambda(\ell) = \lambda e^{\epsilon
\ell} \nonumber\\
r(\omega;\ell) &=& -c_1 K(\omega/\omega_0)\lambda(\ell)/\epsilon
\nonumber\\
\lambda(\omega,\omega';\ell)&=&  -c_2 K(\omega/\omega_0)
K(\omega/\omega_0) \lambda^2(\ell)/(2\epsilon) \label{rgrel}
\end{eqnarray}
Clearly, Eq.\ \ref{rgrel} is valid till $\lambda(\ell) \simeq 1$
after which the system flows towards the strong-coupling fixed point
and the perturbative RG does not hold an more. This happens at
$\ell_3= \ln(1/\lambda)/\epsilon$. Thus we analyze the regime where
$\ell_2 \le \ell_1,\ell_3$ which requires the frequency to satisfy
\begin{eqnarray}
\omega_0 \ge r^{(z+1)/2}, \lambda^{(z+1)/\epsilon}. \label{condrel}
\end{eqnarray}
If the drive frequency satisfies Eq.\ \ref{condrel}, RG stops at
$\ell=\ell_2$ and in this regime the condition of non-Gaussian
behavior is given by
\begin{eqnarray}
\lambda/r \ge \omega_0^{(2-d-z)/(z+1)}.  \label{ngrel}
\end{eqnarray}
Eq.\ \ref{ngrel} shows that the onset of the non-gaussian regime
occurs in a qualitatively different manner for $d+z<2$ since here
$\lambda(\ell)$ grows faster than $r(\ell)$. For smaller $\omega_0$,
where the RG flow stops at larger RG time $\ell_2$,
$\lambda(\ell_2)/r(\ell_2)$ may become large even for smaller
initial value $\lambda/r$ and lead to the onset of the non-gaussian
regime. Thus for any given $\lambda/r$ there exists a upper critical
frequency $\omega_0^{u} \simeq (\lambda/r)^{(z+1)/(2-d-z)}$ below
which the system sees the onset of the non-gaussian regime. In
contrast for $d+z>2$, where the interaction either grows slower than
the mass term or decays, one needs a finite drive frequency greater
than a lower critical frequency $\omega_0^{l} \simeq
(r/\lambda)^{(z+1)/|2-d-z|}$ to achieve the non-gaussian regime. For
$z+d=2$, the onset of the non-gaussian regime requires $\lambda>r$
since both $r(\ell)$ and $\lambda(\ell)$ scales in the same way. We
note however, that at higher loops in RG this relation is expected
to be modified due to the presence of a non-zero anomalous exponent
$\eta$; this point is discussed in details in Sec.\ \ref{diss}.

The analysis of the RG equations for $\epsilon>0$ and $K(0) \ne 0$
turns out to be more complicated and we do not attempt it here.

\section{Analysis of equations of motion}
\label{gpan}

In this section, we shall derive the RG equations from an equation
of motion approach. Although, the end results are the same, we carry
out this analysis to establish a connection between these two
approaches; the latter being widely used in the statistical
mechanics community for studying classical non-equilibrium
phenomenon. Our approach here will be along the same lines as Ref.\
\onlinecite{anatoly2}.

The saddle point equations of the $\phi^4$ action is obtained by
$\delta S/\delta\pca(\kk,k_0)=0=\delta S/\delta\pqa(\kk,k_0)$ which
yields the equation of motion for the fields
\begin{widetext}
\bqa
G_0^{-1}(\kk,k_0)\pq(\kk,k_0)&=&\frac{\lambda}{\omega_0}\prod_{i=2}^3\prod_{j=2}^4\int\frac{d^d\kk_id\omega_j}{(2\pi)^{2d+3}}
\,K(\frac{\omega}{\omega_0})
\Big[\pqa(\kk_2,\omega_2)\pc(\kk_3,\omega_3)\pc(\kk+\kk_2-\kk_3,\omega_4)\no\\&&
+\pqa(\kk_2,\omega_2)\pq(\kk_3,\omega_3)\pq(\kk+\kk_2-\kk_3,\omega_4)
+2\pca(\kk_2,\omega_2)\pc(\kk_3,\omega_3)\pq(\kk+\kk_2-\kk_3,\omega_4)\Big]\no\\
G_0^{-1}(\kk,k_0)\pc(\kk,k_0)&=&\frac{\lambda}{\omega_0}\prod_{i=2}^3\prod_{j=2}^4\int\frac{d^d\kk_id\omega_j}{(2\pi)^{2d+3}}
\,K(\frac{\omega}{\omega_0})
\Big[\pca(\kk_2,\omega_2)\pc(\kk_3,\omega_3)\pc(\kk+\kk_2-\kk_3,\omega_4)\no\\&&
+\pca(\kk_2,\omega_2)\pq(\kk_3,\omega_3)\pq(\kk+\kk_2-\kk_3,\omega_4)
+2\pqa(\kk_2,\omega_2)\pq(\kk_3,\omega_3)\pc(\kk+\kk_2-\kk_3,\omega_4)\Big],\no\\
\label{eom1} \eqa\end{widetext} where
$\omega=k_0+\omega_2-\omega_3-\omega_4$.

Next, we carry lower the momentum cut-off from $\la$ to $\la
e^{-l}$. To this end, we separate the field into slow and fast modes
$\phi(\kk,k_0)=\phl(\kk,k_0)
\Theta(\la-|\kk|)+\phg(\kk,k_0)\Theta(|\kk|-\la)$. Using Eq.\
\ref{eom1}, we write down the equations for $\phg(\kk,k_0)$ and find
the propagator for the fast modes. In doing so, we make the
following approximations. First, we retain only part of the
interaction term for which $\Theta(|\kk|-\la)$ are satisfied.
Second, we ignore the terms in the right side of Eq.\ \ref{eom1}
which has more than one $\phg(\kk,k_0)$. This approximation is
equivalent to replacing the full $\mathcal{G}^>$ in the action
formalism by the free propagator $G^>_0$ which is the standard
approximation in perturbative RG procedure. This leaves out terms
with one $\phg$ and two $\phl$ which constitutes the self energy of
$\phg$ fields due to the interaction between the fast ($\phg$) and
the slow ($\phl$) field modes. This yields \bqa [G_0^{-1}(k)
\delta_{\alpha \beta} \delta(k-k') -\Sigma_{\alpha \beta}(k,k')]
\phg_{\beta}(k') &=& 0, \label{eom2}\eqa where $k\equiv (\kk,k_0),
k^\prime\equiv (\kk^\prime,k^\prime_0)$, the indices $\alpha$ and
$\beta$ takes values $1(2)$ for $c(q)$, all repeated indices are
summed or integrated over, and $\phg \equiv (\phg_{c},\phg_q)$. The
self-energy $\Sigma_{\alpha \beta}$ satisfies
$\Sigma_{11}=\Sigma_{22} \equiv \Sigma_1$ and
$\Sigma_{12}=\Sigma_{21}\equiv \Sigma_2$ which are given by \bqa
\Sigma_1(k_1,k_2)&=&-\frac{2\lambda}{\omega_0}\int^\prime\frac{d^d\kk
dk_0
d\omega}{(2\pi)^{d+2}}K(\frac{\omega}{\omega_0})\Big[\phi^<_c(\kk,k_0)\no\\&&
\times \phi_c^<(\kk+\kk_1-\kk_2,k_0+\omega_1-\omega_2-\omega)\no\\&&+c\leftrightarrow q\Big]\no\\
\Sigma_2(k_1,k_2)&=&-\frac{2\lambda}{\omega_0}\int^\prime\frac{d^d\kk
dk_0
d\omega}{(2\pi)^{d+2}}K(\frac{\omega}{\omega_0})\Big[\phi^<_c(\kk,k_0)\no\\&&
\times
\pql(\kk+\kk_1-\kk_2,k_0+\omega_1-\omega_2-\omega)\no\\&&+\mbox{h.c}\Big]
\eqa The Keldysh Green function for the $\phg$ fields can thus be
obtained as
\begin{widetext}
\bqa \mathcal{G}^>(k_1,k_2)&=&\left(\begin{array}{cc}
\mathcal{G}_K(k_1,k_2) & \mathcal{G}_R(k_1,k_2)\\
\mathcal{G}_A(k_1,k_2) & 0
\end{array}
\right)\no\\
\mathcal{G}_K(k_1,k_2)&=&
G_K(k_1)\delta(k_1-k_2)+G_K(k_1)\Sigma_1(k_1,k_2)G_K(k_2)
+G_K(k_1)\Sigma_2(k_1,k_2)G_A(k_2)\no\\&&
+G_R(k_1)\Sigma_2(k_1,k_2)G_K(k_2)
+G_R(k_1)\Sigma_1(k_1,k_2)G_A(k_2)\no\\
\mathcal{G}_R(k_1,k_2)&=&G_R(k_1)\delta(k_1-k_2)+G_K(k_1)\Sigma_1(k_1,k_2)G_R(k_2)+G_R(k_1)\Sigma_2(k_1,k_2)G_R(k_2)\no\\
\mathcal{G}_A(k_1,k_2)&=&G_A(k_1)\delta(k_1-k_2)+G_A(k_1)\Sigma_1(k_1,k_2)G_K(k_2)+G_A(k_1)\Sigma_2(k_1,k_2)G_A(k_2)
\label{eom3}\eqa
\end{widetext}

Next, we write down the equation for the slow-modes from Eq.\
\ref{eom1} and average out the fast modes from that equation by
replacing the $\phga_\kk\phg_{\kk^\prime}$ by their average from
Eq.\ \ref{eom3}. After some straightforward algebra, one obtains
\begin{widetext}\bqa
G_0^{-1}(k)\pql(k)&=&\frac{2\lambda}{\omega_0}\int\frac{d^d\kk_3d\omega_3
d\omega}{(2\pi)^{d+2}}K(\frac{\omega}{\omega_0})
\Big\{\pql(\kk_3,\omega_3-\omega)\int^\prime\frac{d^d\kk_2d\omega_2}{(2\pi)^{d+1}}
\mathcal{G}_K(k_2,k_2+k-k_3) \no\\&&
+\phi_c^<(\kk_3,\omega_3-\omega)\int^\prime\frac{d^d\kk_2d\omega_2}{(2\pi)^{d+1}}\Big[\mathcal{G}_A(k_2,k_2+k-k_3)
+\mathcal{G}_R(k_2,k_2+k-k_3)\Big]\Big\}\no\\
&&+ \frac{\lambda}{\omega_0}\prod_{i=2}^3\prod_{j=2}^4
\int\frac{d^d{\kk}_id\omega_j}{(2\pi)^{2d+3}}
\,K(\frac{\omega}{\omega_0})
\Big[\pqla(\kk_2,\omega_2)\phi_c^<(\kk_3,\omega_3)\pcl(\kk+\kk_2-\kk_3,\omega_4)\no\\&&
+\pqla(\kk_2,\omega_2)\pql(\kk_3,\omega_3)\pql(\kk+\kk_2-\kk_3,\omega_4)
+2\phi_c^<(\kk_2,\omega_2)\phi_c^<(\kk_3,\omega_3)\pql(\kk+\kk_2-\kk_3,\omega_4)\Big]\label{eom4}
\eqa\end{widetext} The equation for $\phi^{<}_{c}(k)$ is obtained by
$\phi^{<}_c(k) \leftrightarrow\phi_q^{<}(k)$ in the above equation.

The additional term arising from replacing the $\phg$ fields with
their averages is the first of the two terms in the right side of
Eq.\ \ref{eom4}. Substituting $G$ from Eq.\ \ref{eom3}, we find that
to $\mathcal{O}(\lambda)$, the additional term in the equation of
motion for the $\phl$ fields is given by \bqa
G_0^{-1}(k)\phl(k)&=&\frac{2\lambda}{\omega_0}\int\frac{d^d\kk_3d\omega_3
d\omega}{(2\pi)^{d+2}}K(\frac{\omega}{\omega_0})\no\\&&
\sigma_x\phl(\kk_3,\omega_3-\omega)Tr[G_K] + ...,  \eqa where $
Tr[G_K]=\int^\prime\frac{d^d\kk dk_0}{(2\pi)^{d+1}}G_K(\kk,k_0)$ and
the ellipsis denote all other terms in the right side of Eq.\
\ref{eom4} and its counterpart for  $\phi^{<}_{c}(k)$ which is
obtained by substituting $\phi^{<}_c(k)
\leftrightarrow\phi_q^{<}(k)$ in Eq.\ \ref{eom4}. Thus we find that
in exact accordance with the results obtained by implementing RG on
the action, the $\omega=0$ part of this term provides the one loop
correction to the $r$ while the $\omega\ne 0$ part generates new
terms in the equation of motion which are same as those obtained
from Eq.\ \ref{intac3}.

A similar result for the one-loop corrections to
$\mathcal{O}(\lambda^2)$ terms can be obtained by gathering the
terms originating from $G$ to $O(\lambda^2)$. For example, the terms
which contribute to the correction of the term $\pcla\pcl\pcl$ are
given by
\begin{widetext}
\bqa
\left(\frac{2\lambda}{\omega_0}\right)^2\int\frac{d^d\kk_1d^d\kk_3d\omega_1d\omega_3d\omega
d\omega^\prime}{(2\pi)^{2d+4}}
K(\frac{\omega}{\omega_0})K(\frac{\omega^\prime}{\omega_0})
\phi_c^<(\kk_1,\omega_1)\phi_c^<(\kk_3,\omega_3-\omega)\phi_c^<(\kk_1+\kk_3-\kk,\omega_1+\omega_3-k_0-\omega^\prime)\no\\
\int^\prime\frac{d^d\kk_2d\omega_2}{(2\pi)^{d+1}}\Big[
G_A(k_2)G_K(k_2+k-k_3)+G_K(k_2)G_R(k_2+k-k_3)+G_K(k_2)G_K(k_2+k-k_3)\no\\+G_R(k_2)G_A(k_2+k-k_3)+G_A(k_2)G_R(k_2+k-k_3)\Big]\no
\eqa\end{widetext} The last three terms cancel in the limit of zero
external frequency and momenta. The rest of the terms provide the
same loop corrections to $\lambda$ and generate the new terms in the
equation of motion as those obtained from Eq.\ \ref{intac3}.

Finally, we complete the RG procedure by scaling the momenta and
frequency by $k \to k e^{\ell}$ and $\omega \to \omega e^{z\ell}$.
The scaling of the fields are the same as those obtained in Sec.\
\ref{keldyshsec} and leads to $\omega_0 \to \omega_0 e^{z\ell}$ for
the drive frequency. Gathering the RG terms generated from the
scaling and the one-loop corrections described above, we finally
obtain Eqs.\ \ref{rgeq1} and \ref{intac2}.

\section{Discussion}
\label{diss}

In this work, we have aimed at providing a perturbative RG approach
to understanding the properties of a driven quantum system. We have
illustrated the main points of our work by deriving and analyzing
the RG equations for a system described by a scalar bosonic $\lambda
\phi^4$ theory. There are numerous concrete examples of such
effective field theories in condensed matter physics; several
quantum models (such as the Bose Hubbard and the Ising models) near
their critical point is described by such a field theory. In fact,
almost all quantum critical systems which are described by a
Landau-Ginzburg action of a single component order parameter field
admits an analogous description in their ordered phase near the
quantum critical point. We expect our RG analysis to hold for such
systems.

The key results that emerge from our analysis are the following.
First we show that the drive frequency scales like the physical
temperature in equilibrium systems and sets the cutoff scale for RG
given by $\ell_2= \ln(\Lambda v_0/\omega_0)/(z+1)$. Second, when the
drive frequency $\omega_0$ and the effective mass term $r_{\rm eff}$
satisfies $\omega_0/r^{2/z} \ll1$, the drive can be treated
perturbatively and one expect the universality of such a system to
be analogous to its equilibrium counterpart. In this regime, the RG
flow stops when the cutoff reaches the system correlation length at
$\ell=\ell_1 \le \ell_2$ and the drive do not qualitatively alter
the behavior of the flow. Third, we show that in the other regime
where $\ell_2 \le \ell_1$ which occurs when $r_{\rm eff} \le
\omega_0^{2/(z+1)}$, the drive dominates the physics and may lead to
setting in of a non-gaussian regime. For drive protocols with
$K(0)=0$ [or $\alpha_0=0$ for periodic protocols], the condition for
setting in such a regime is $\lambda/r \ge \omega_0^{(2-d-z)/(z+1)}$
for any $d+z$. This relation clearly distinguishes between the
behaviors of systems with $d+z<2$ and $d+z>2$. For the former, there
exists a upper critical drive frequency below which the non-gaussian
regime sets in while for the latter such a setting in occurs above a
lower critical drive frequency. For drive protocols with
$K(0),\alpha_0 \ne 0$, we have shown that the analogous condition
for irrelevant or marginal interaction $d+z\le 4$ is
$\lambda'/r_{\rm eff} \ge \omega_0^{(2-d-z)/(z+1)}$. Finally, we
note that the present scheme can be easily generalized to drive
protocols with multiple frequencies; for those drives, the highest
characteristic frequency scale assumes the role of $\omega_0$.

We also note that the setting in of the non-gaussian regime occurs
concomitantly with $r(\omega)$ (or $r_n$ for periodic drive)
becoming comparable with $r$. This indicates that the effective
Hamiltonian which describes the dynamics in this regime will have
non-perturbative mode coupling terms. Consequently, one expects that
the energy pumped in the system due to the drive will be efficiently
distributed between the different modes. Thus the system can
effectively absorb large amounts of energy at long time. In
contrast, in the gaussian regime, the mode coupling terms can be
treated perturbatively and the presence of a large mass gap prevents
the system to have large excess energy. The crossover between the
two regimes has been argued in Ref.\ \onlinecite{anatoly2}, using a
Magnus expansion approach for a one-dimensional spin chain, to be
the signature of a energy localization-delocalization transition.
Our RG analysis shows a similar behavior and provides a criterion
for such a crossover to occur; however, deciphering the precise
relation of the present general analysis with the specific
quantitative study of Ref.\ \onlinecite{anatoly2} would require
further study. We also note that for the present system such a
crossover is not expected to occur if the drive term involved a
time-dependent $r$; in this case the mode coupling terms are present
in the starting Hamiltonian and grow under RG. Consequently, the
system is expected to continue to absorb energy indefinitely and
hence be always delocalized in energy.

Another generalization of our work would involve working out the RG
equations to two loops. One expects such a calculation to unravel
the dependence of the condition of setting in of the non-Gaussian
regime on the anomalous dimension $\eta$. The lowest-order non-zero
contribution to $\eta$ comes from two-loop RG diagrams and hence
such a dependence can not be studied within the one-loop RG analysis
carried out here. The simplest guess to the nature of such a
correction is as follows. The scaling dimension of the fields
$\phi({\bf k},\omega)$, in the presence of finite $\eta$, is given
by $\alpha'=(2+d+z+\eta)/2$. Using this, a straightforward power
counting shows that $ \epsilon \to \epsilon'= 4-d-z+2\eta$. This
means that $\lambda(\ell) \sim \lambda e^{\epsilon' \ell}$ (for all
protocols with $K(0)$ or $\alpha_0=0$) and $r \sim r^{2 \ell}$. Thus
at a scale $\ell=\ell_2$, the condition for the non-Gaussian
behavior would be modified to $ \lambda/r \ge
\omega_0^{(2-d-z+2\eta)/(z+1)}$. Note that this is extremely
important for systems with $d+z=2$ where $\eta$ is expected to
provide the entire frequency dependence. However, this guess needs
to be substantiated with full two-loop RG calculations which is left
as a topic for future study.

Finally, we note that there the present RG technique allows for
several other extensions. First, it will be interesting to study the
consequence of driving an open quantum system in the presence of a
bath at a finite temperature which would allow for noise and
dissipation using this scheme. Such a study has recently been
carried out using functional RG in Ref.\ \onlinecite{ehud1};
however, their work did not involve a time-dependent drive protocol
which is the main focus of the present study.  Second, the RG
procedure could be easily generalized to actions describing bosonic
fields with $N>1$ components. Finally, it would be interesting to
carry out a similar RG for fermions where the presence of a Fermi
surface is expected to provide new features for the RG flow. We plan
to undertake these studies in future.

\section{Appendix: RG calculation}
\label{appa}

In this section, we provide a detailed derivations of the RG
equations. Our analysis will be primarily carried out for theories
with $z=2$, but similar results can be obtained for $z=1,3$. From
Eq.\ \ref{ac1}, we note that $S_0$ represents the quadratic part of
the action and leads to the Green functions given by \bqa
G_K({\bf k},\omega)&=&\langle \pcla({\bf k},\omega)\pcl({\bf k} ,\omega)\rangle\no\\
&=&[1+2n_B(\omega)]\delta(\omega-E_{\bf k}) \nonumber\\
G_{R}({\bf k},\omega)&=&\langle \pqa({\bf k},\omega)\pcl({\bf k} ,\omega)\rangle\no\\
&=& \frac{1}{\omega-E_{\bf k}+i\eta} = G_A^{\ast}({\bf k},\omega).
\label{kgf1} \eqa The interaction term $S'_1$, in frequency and
momentum space is given by Eq.\ \ref{intac1}.

\begin{figure}[t!]
\rotatebox{0}{
\includegraphics*[width=0.6\linewidth]{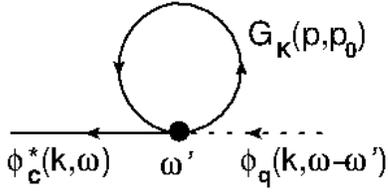}}
\caption{Tadpole diagram for one-loop correction to mass term.}
\label{rcorr}
\end{figure}

\begin{figure}[t!]
\includegraphics*[width=\linewidth]{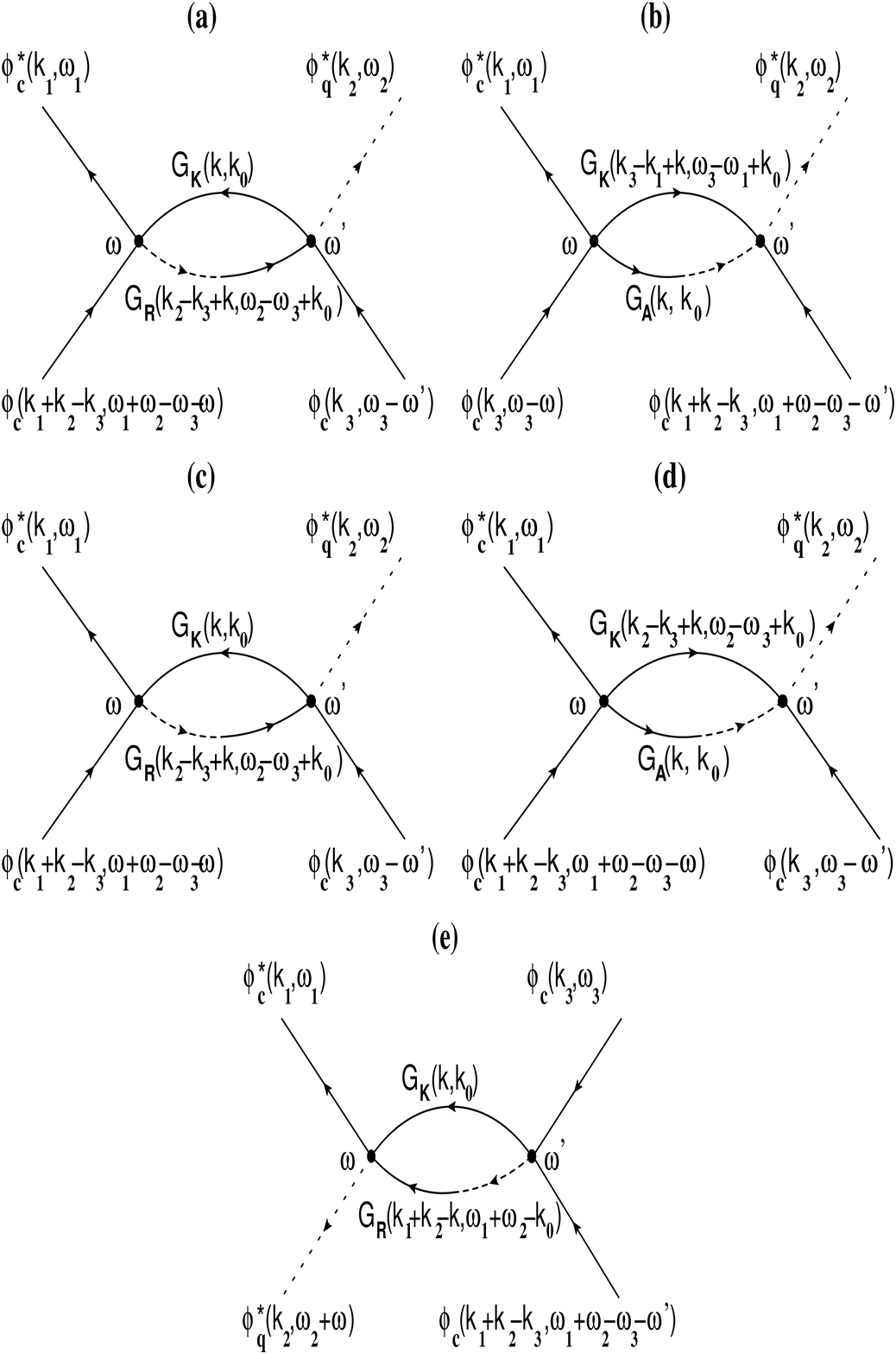}
\caption{1-loop correction to the interaction term in the effective
action} \label{lcorr}
\end{figure}

Here we consider the perturbative corrections that originate from
integrating out the field modes. To linear order in $\lambda$ such a
term is given by the diagram shown in Fig.\ \ref{rcorr} which leads
to the one-loop correction to $r$. One such term is given by \bqa
\delta S_2 &=& -4\lambda\int \frac{d^dk_1..d^dk_3
d\omega_1..d\omega_4}{(2\pi)^{3d+4}\omega_0}\, \phi^{\ast
<}_{c}({\bf k}_1,\omega_1)\phi_{q}^<({\bf k}_2,\omega_2)\no\\&&
\times \langle\phi^{\ast >}_{c}({\bf
k}_3,\omega_3)K(\frac{\omega_1-\omega_2+\omega_3-\omega_4}{\omega_0})
\nonumber\\
&& \times \phi_{c}^>({\bf k}_1-{\bf
k}_2+{\bf k}_3,\omega_4)\rangle_{S_{0>}}\no\\
&=& -4\lambda \int \frac{d^dk
d\omega_1d\omega_2}{(2\pi)^{d+2}\omega_0}\,K(\frac{\omega_1-\omega_2}{\omega_0})
\no\\&& \phi^{\ast <}_{c}({\bf k},\omega_1)\phi_{q}^<({\bf
k},\omega_2)\mbox{Tr}[G_K]. \eqa Other terms can be obtained in a
similar fashion and finally we obtain \bqa \delta S_2&=&-4\lambda
\int \frac{d^dk
d\omega_1d\omega_2}{(2\pi)^{d+2}\omega_0}\,K(\frac{\omega_1-\omega_2}{\omega_0})
\no\\&& \phi^{\ast <}({\bf k},\omega_1)\sigma_x\phi^<({\bf
k},\omega_2)\mbox{Tr}[G_K] \eqa
To make further progress we divide the terms into a piece for which
$\omega_1=\omega_2$ contributing to the renormalization of $r$ and
other terms for which $\omega_1\ne \omega_2$. This is formally done
by writing \bqa K(\omega/\omega_0)&=&K(\omega/\omega_0)[
\delta(\omega/\omega_0) + (1-\delta(\omega/\omega_0)]
\label{ksplit}. \eqa
The first terms yields the one loop correction to $r$ in the action
\bqa -4\lambda\int \frac{d^dk d\omega}{(2\pi)^{d+1}} K(0)\phi^{\ast
<}({\bf k},\omega)
\sigma_x\phi^<({\bf k},\omega)\mbox{Tr}[G_K]\no\\
\eqa leading to the RG equation for $r$ \bqa
\frac{dr(\ell)}{d\ell}&=&2r+c_1(\la) K(0) \lambda(\ell), \eqa where
$C_1=4\Gamma_d \la^{d-1} [1+2n_B(E_\la)]$, and $\Gamma_d$ is the
angular integral in $d$ dimensions. The second term generates new
coupling terms in the action which is of the form \bqa \delta S_1
&=& -\int \frac{d^dk d\omega d\omega_1}{(2\pi)^{d+1} \omega_0}
 \phi^{\ast <}({\bf k},\omega_1)r(\omega) \sigma_x \nonumber\\
 && \times \phi^<({\bf k},\omega_1-\omega) [1-\delta(\omega/\omega_0)]
\eqa with the RG equation for $r(\omega) \equiv r(\omega,\ell)$
given by \bqa \frac{dr(\omega,\ell)}{d\ell}&=& c_1(\la)
K(\omega/\omega_0) \lambda(\ell).  \eqa

Next, we consider the one-loop correction to the quartic coupling
$\lambda$. The diagrams which contribute to the 1-loop correction to
the term $\phi_c^{\ast <}\phi_q^{\ast <}\pc\pc$ are shown in the
Fig.\ref{lcorr}. The first two diagrams [(a)+(b)] evaluate to \bqa
\delta S_4&=&-4\lambda^2\int \frac{d^dk_1..d^dk_3
d\omega_1..d\omega_3d\omega^\prime
d\omega}{(2\pi)^{3d+5}\omega^2_0}\,
K(\frac{\omega^\prime}{\omega_0}) \no\\&& \times
K(\frac{\omega}{\omega_0}) \times \mbox{Tr}[G_K
G_R+G_A G_K]_{{\bf k}_i=0} \nonumber\\
&& \times \phi_c^{\ast <}({\bf k}_1,\omega_1)\phi_q^{\ast <}({\bf
k}_2,\omega_2) \phi_c^<({\bf k}_3,\omega_3-\omega)\no\\&& \times
\phi_c^<({\bf k}_1-{\bf k}_2+{\bf k}_3,
\omega_1+\omega_2-\omega_3-\omega^\prime) \eqa where Tr$[G_K G_R+G_A
G_K]$ is \bqa \int_{\la-d\la}^\la\frac{d^dk dk_0}{(2\pi)^{d+1}}
[G_K(k,k_0)G_R(k_3-k_1+k,\omega_3-\omega_1+k_0)
\no\\+G_A(k,k_0)G_K(k_3-k_1+k,\omega_3-\omega_1+k_0)]\no\\
= \Gamma_d
\la^{d-1}d\la\frac{[1+2n_B(E_\la)]-[1+2n_B(E_{k_3-k_1+\la})]}
{\omega_3-\omega_1+E_\la-E_{k_3-k_1+\la}} \no\ \ \eqa Taking the
limits of zero external frequencies and momenta, one gets \bqa
\mbox{Tr}[G_K G_R+G_A G_K]|_{{\bf k}_i,\omega_i=0} &=& 2\Gamma_d
\la^{d-1}d\la \frac{\partial n_B}{\partial E}. \label{rgev} \eqa The
other terms can be evaluated in a similar manner. Once again, we
find that one can split $K(\omega/\omega_0)$ in to $\omega=0$ and
$\omega \ne 0$ parts using Eq.\ \ref{ksplit}. The former provides
correction to the $\lambda$ term whereas the latter generates new
terms in the action. The correction to the $\lambda$ coupling is
given by \bqa \delta S_4^1&=&4\lambda^2 K(0)\int
\frac{d^dk_1..d^dk_3
d\omega_1..d\omega_3d\omega^\prime}{(2\pi)^{3d+4}\omega_0}
K(\frac{\omega^\prime}{\omega_0})\no\\&& \times \phi^{\ast <}_c({\bf
k}_1,\omega_1) \phi_q^{\ast <}({\bf k}_2,\omega_2)\phi_c^<({\bf
k}_3,\omega_3)\no\\&& \times \phi_c^<({\bf k}_1-{\bf k}_2+{\bf
k}_3,\omega_1+\omega_2-\omega_3-\omega^\prime) \no\\&& \times
\mbox{Tr}[G_K G_R+G_A G_K]_{{\bf k}_i=0} \eqa leading to RG equation
for $\lambda$ \bqa
\frac{d\lambda(\ell)}{dl}&=&\ep\lambda(\ell)-c_2(\la)K(0)\lambda^2(\ell)
\eqa where $c_2(\la)d\la=-4 \mbox{Tr}[G_K G_R+G_A G_K]_{{\bf
k}_i=0}$. The expression for $c_2(\Lambda)$ can thus be directly
read off from Eq.\ \ref{rgev}.

The new terms in the action are of the form \bqa \delta S_4^2&=&
\int \frac{d^dk_1..d^dk_3 d\omega_1..d\omega_3d\omega^\prime d
\omega}{(2\pi)^{3d+5}\omega_0}
K\left(\frac{\omega^\prime}{\omega_0}\right) \no\\&& \times
\lambda(\omega,\omega^\prime) \phi_c^{\ast <}({\bf
k}_1,\omega_1)\phi_q^{\ast <}({\bf k}_2,\omega_2)\phi_c^<({\bf
k}_3,\omega_3-\omega) \nonumber\\ && \times \phi_c^<({\bf k}_1-{\bf
k}_2+{\bf k}_3,\omega_1+\omega_2-\omega_3-\omega^\prime) \eqa with
the RG equations  for $\lambda(\omega,\omega) \equiv
\lambda(\omega,\omega';\ell)$ given by
\bqa \frac{d\lambda(\omega,\omega^\prime,\ell)}{d\ell}&=&
-c_2(\la)K(\omega^\prime/\omega_0)K(\omega/\omega_0)\lambda^2(\ell).
\eqa

Finally, we note that when the drive is periodic, the function
$K(\omega/\omega_0)= \sum_{n} \alpha_n \delta(n-\omega/\omega_0)$
has support only on a set of discrete points. In this case, it is
easier to write $ K(\omega/\omega_0)=\sum_n \alpha_n/\pi \lim_{\eta
\to 0} \eta/[\eta^2 +(\omega/\omega_0-n)^2]$. The analysis for this
form of $K$ is then easily carried out and obtains Eqs.\ \ref{rgeq2}
with the identification $K(\omega/\omega_0) \to \alpha_n$. This
completes the derivation of the RG equations used in Sec.\
\ref{keldyshsec}.

Before ending this section, we would like to note that since $r_n$
and $\lambda_{mn}$ are spontaneously generated by the RG flow, one
expects to include this term in the effective action as customary in
the usual RG procedure. We have checked that at least for a simple
drive protocol such as $f(\omega_0 t) = \exp(i \omega_0 t)$, this
leads to additional contribution to the loop diagrams shown in Fig.\
\ref{rcorr} and \ref{lcorr} which are ${\rm O}(r_n^2/\Lambda^2)$ and
${\rm O}(\omega_0^2/\Lambda^2)$ and can thus be ignored. We expect
this feature to hold for other protocols as well.


\begin{thebibliography} {99}
\bibitem{comment1} In principle, there is another scale given by
$\ell'_2 = \ln(T_0)/(z+1)$; however for $\omega_0 \ge T_0^{-1}$,
$\ell'_2 \ge \ell_2$ and thus the RG flow is always cutoff before
this scale is reached.
\end{thebibliography}
\end{document}